\documentstyle[12pt]{article}
\makeatletter
\@addtoreset{equation}{section}
\makeatother

\begin{document}
\begin{center}
{\Large \bf
Conservative Model of Black Hole\\[0.5cm]
and Lifting of the Information Loss Paradox}
\\[1.5cm]
{\bf Vladimir S.~MASHKEVICH}\footnote {Email:
mashkevich@gluk.apc.org}  \\[1.4cm]
{\it Institute of Physics, National academy
of sciences of Ukraine \\
252028 Kiev, Ukraine} \\[1.4cm]
\vskip 1cm

{\large \bf Abstract}
\end{center}

The conservative model of a black hole is advanced.
The model incorporates conservation laws such as those
of baryon and lepton numbers, which lifts the information
loss paradox. A scenario of black hole evaporation is
considered.

\newpage

\section*{Introduction}

Over twenty years ago, Steven Hawking made a striking and
seminal discovery that basic principles of quantum field
theory lead to the emission of thermal radiation from a
classical black hole [1,2]. This discovery gave rise to a
famous paradox---the information loss paradox of black
hole physics (see, e.g., [3-7]). The paradox has several
aspects. In this paper, we point out the following of them:

(i) the violation of unitary time evolution;

(ii) a startlingly great value of black hole entropy;

(iii) the problem of accommodating the entropy after black
hole evaporation;

(iv) the violation of conservation laws.

As for the aspect (i), an issue relating to it is by no means
new for physics: The issue is the problem of quantum
indeterminism, i.e., quantum jumps resulting in information
loss. The discussion of this problem is outside this paper.

The aspect (ii) is that no ordinary physical system has as
great entropy as that attributed to a black hole.

The aspect (iii) lies in the fact that the whole of the
universe is too small to accommodate the entropy of the
black hole after its evaporation.

It is the aspect (iv) that plays a pivotal role in lifting
the information loss paradox. The violation of the conservation
laws such as those of baryon and lepton numbers is implicitly
put into the determination of the rate of the Hawking emission.
The violation results in that the only channel of black hole
energy loss is that through heat. It is this fact that leads
to the immense value of the black hole entropy.

The situation reverses in a cardinal way if from the outset
we require the conservation laws be respected. Then there is
another channel of energy loss---that through the decrease
of the number of particles conserved. This results in a
possibility for a normal value of the black hole entropy,
which lifts the paradox.

A model introduced in this paper may be called the conservative
model of a black hole: The model incorporates conservation
laws and is as physically conservative as possible.

\section{The information loss paradox}

A conventional treatment of the Hawking emission is as follows.
Occupation numbers relating to the emission are
\begin{equation}
n=\frac{1}{e^{\omega/T}\mp 1},
\label{1.1}
\end{equation}
the minus (respectively plus) sign corresponding to a bosonic
(respectively fermionic) mode with the frequency $\omega$,
where the Hawking temperature $T$ is
\begin{equation}
T=\frac{1}{8\pi\kappa E},
\label{1.2}
\end{equation}
$E=M$ is black hole energy, and $\kappa$ is the gravitational
constant.

The black hole horizon radius $R_{S}$ is
\begin{equation}
R_{S}=2\kappa M
\label{1.3}
\end{equation}
and the relating area $A$ is
\begin{equation}
A=4\pi R_{S}^{2}\;,
\label{1.4}
\end{equation}
so that we have an identity
\begin{equation}
dE=\frac{1}{8\pi\kappa E}d\left( \frac{A}{4\kappa} \right).
\label{1.5}
\end{equation}

On the other hand, eq.(\ref{1.1}) implies that the only
channel of black hole energy loss is that through heat,
i.e.,
\begin{equation}
dE=\delta Q.
\label{1.6}
\end{equation}
For an equilibrium process,
\begin{equation}
\delta Q=TdS
\label{1.7}
\end{equation}
holds where $S$ is entropy, so that we obtain
\begin{equation}
dE=TdS.
\label{1.8}
\end{equation}
Eqs.(\ref{1.5}),(\ref{1.8}),(\ref{1.2}) give
\begin{equation}
dS=d\left( \frac{A}{4\kappa} \right).
\label{1.9}
\end{equation}
It follows from eqs.(\ref{1.4}),(\ref{1.3}) that
\begin{equation}
A=0\quad {\rm for}\quad M=0.
\label{1.10}
\end{equation}
Assuming that
\begin{equation}
S=0\quad {\rm for}\quad M=0
\label{1.11}
\end{equation}
takes place, we obtain the equation
\begin{equation}
S=\frac{A}{4\kappa}.
\label{1.12}
\end{equation}
Indeed, Bekenstein proposed [8] that
\begin{equation}
S\approx \frac{A}{4\kappa}.
\label{1.13}
\end{equation}
It is eq.(\ref{1.12}) that is the primary source of the
information loss paradox.

We have from eqs.(\ref{1.12}),(\ref{1.4}),(\ref{1.3})
\begin{equation}
S\approx \left( \frac{M}{m_{P}} \right)^{2}
\label{1.14}
\end{equation}
where $m_{P}$ is the Planck mass, so that for the solar
mass $M$ we obtain
\begin{equation}
S\approx \left( \frac{10^{30}}{10^{-8}} \right)^{2}=
10^{76}
\label{1.15}
\end{equation}
---an enormous value.

Let us ascribe the entropy (\ref{1.14}) to a thermal
radiation, for which
\begin{equation}
M=E=4\sigma V_{tr}T_{tr}^{4},\quad S=\frac{16}{3}\sigma
V_{tr}T_{tr}^{3}
\label{1.16}
\end{equation}
where $\sigma$ is the Stefan-Boltzmann constant, $V$ is
the volume, and $tr$ stands for thermal radiation. We find
for the solar mass $M$
\begin{equation}
V_{tr}\approx 10^{86}m^{3}>>>10^{78}m^{3}\approx V_{universe},
\label{1.17}
\end{equation}
so that the universe is too small to accommodate the entropy of
the black hole after its evaporation. Thus the information loss
resulting in the entropy (\ref{1.14}) is too great to be
interpreted within the framework of current physics.

\section{Invocation to conservation laws and lifting
of the paradox}

Our aim is to treat a black hole---to the extent which is
possible---as an ordinary physical system. To put this
another way, we want to be as conservative as possible.
Thus far, we face two nonordinary features. The first
feature is eq.(\ref{1.12}) resulting in an enormous value
of the black hole entropy. The second feature is as follows.
Eq.(\ref{1.1}) for occupation numbers implies that for any
mode a chemical potential $\mu$ equals zero,
\begin{equation}
\mu=0,
\label{2.1}
\end{equation}
which in turn implies that such conservation laws as those of
baryon and lepton numbers are violated. Thus to remedy the
trouble we should remove the two features. Fortunately, this
may be done at one stroke. It suffices from the outset to
require that the conservation laws be respected and to introduce
relating chemical potentials. Then in place of eq.(\ref{1.6}) we
have
\begin{equation}
dE=\delta Q+\sum_{\alpha}\mu_{\alpha}dN_{\alpha}
\label{2.2}
\end{equation}
where $N_{\alpha}$ is the number of particles in the black hole,
so that for an equilibrium process
\begin{equation}
dE=TdS+\sum_{\alpha}\mu_{\alpha}dN_{\alpha}
\label{2.3}
\end{equation}
holds in place of eq.(\ref{1.8}). Eq.(\ref{2.3}) lifts the
paradox.

\section{Black hole as a thermodynamic system}

Eq.(\ref{2.3}) implies that a black hole in an equilibrium
state is described by a grand canonical ensemble. We have
for energy
\begin{equation}
E=E(T,\{\mu_{\alpha}\}).
\label{3.1}
\end{equation}
An essential peculiar feature of the black hole is a
constraint on thermodynamic variables $T,\left\{
\mu_{\alpha} \right\}$:
\begin{equation}
E(T,\left\{ \mu_{\alpha} \right\})=\frac{1}{8\pi
\kappa T},
\label{3.2}
\end{equation}
i.e.,
\begin{equation}
T=T(\left\{ \mu_{\alpha} \right\}).
\label{3.3}
\end{equation}

We have from eqs.(\ref{1.5}),(\ref{1.2}),(\ref{2.3})
\begin{equation}
\frac{1}{T}\sum_{\alpha}\mu_{\alpha}dN_{\alpha}=
d\left( \frac{A}{4\kappa}-S \right);
\label{3.4}
\end{equation}
thus $1/T$ is an integrating factor not only for
$\delta Q$ but also for $\sum_{\alpha}\mu_{\alpha}dN_{\alpha}$.
It goes without saying that
\begin{equation}
S\neq\frac{A}{4\kappa}.
\label{3.5}
\end{equation}

Note that in the conventional treatment, which corresponds to
the case
\begin{equation}
\mu_{\alpha}=0,
\label{3.6}
\end{equation}
the equation
\begin{equation}
E(T)=\frac{1}{8\pi\kappa T}
\label{3.7}
\end{equation}
must be an identity not to fix a value of $T$.

It should be particularly emphasized that as long as
conservation laws are fulfilled, the emission rate depends not
only on $T$ but also on $\left\{ \mu_{\alpha} \right\}$. This
is apparent from the consideration of a process with virtual
particles: those entering the black hole should be
accommodated, i.e., annihilated, for which they must have
counterparts.

\section{Maximal temperature}

A maximal value $T_{max}$ of the temperature is defined by
the equation
\begin{equation}
T_{max}=\frac{1}{8\pi\kappa[E(T_{max},\left\{
\mu_{\alpha} \right\})]_{min}}.
\label{4.1}
\end{equation}
We assume that
\begin{equation}
\frac{\partial E(T,\left\{ \mu_{\alpha} \right\})}
{\partial T}>0,\qquad \frac{\partial E(T,
\left\{ \mu_{\alpha} \right\})}{\partial \mu_{\beta}}>0,
\label{4.2}
\end{equation}
which is typical for a grand canonical ensemble. Then
we have
\begin{equation}
[E(T_{max},\left\{ \mu_{\alpha} \right\})]_{min}=
E(T_{max},\left\{ \mu_{\alpha}=-\infty \right\}),
\label{4.3}
\end{equation}
which corresponds to
\begin{equation}
N_{\alpha}=0.
\label{4.4}
\end{equation}
The minimal value of $E$ implies that the state
$(T_{max},\left\{ \mu_{\alpha}=-\infty \right\})$ corresponds
to the presence of a single quantum $\omega$. In view of
eq.(\ref{1.1}), we have
\begin{equation}
\omega\approx T_{max}
\label{4.5}
\end{equation}
and by eq.(\ref{3.2})
\begin{equation}
\omega\approx \frac{1}{\kappa\omega}.
\label{4.6}
\end{equation}
Thus we have for the maximal temperature and the last quantum
\begin{equation}
\omega_{last}\approx T_{max}\approx m_{P}.
\label{4.7}
\end{equation}

\section{A scenario of black hole evaporation}

Let us consider a simple scenario of black hole evaporation.
Let there be only one $\alpha$, so that a set $\left\{
\mu_{\alpha} \right\}$ reduces to the only $\mu$. We consider
a process of evaporation with maximal equilibrium and
maximal confinement of a black hole atmosphere. We have
equations:
\begin{equation}
E+\tilde{E}_{r}=E_{\Sigma}={\rm const},\quad
N+\tilde{N}_{r}={\rm const},
\label{5.1}
\end{equation}
and
\begin{equation}
\tilde{E}_{r}=E_{a}+E_{r},\quad
\tilde{N}_{r}=N_{a}+N_{r}
\label{5.2}
\end{equation}
where $r$ stands for rest and $a$ for atmosphere.

Stages of the scenario are as follows.

A state: total confinement,
\begin{equation}
E(T,\mu)+E_{a}(V_{a},T,\mu)=E_{\Sigma},
\label{5.3a}
\end{equation}
\begin{equation}
N(T,\mu)+N_{a}(V_{a},T,\mu)=N_{\Sigma},
\label{5.3b}
\end{equation}
\begin{equation}
T=\frac{1}{8\pi\kappa E(T,\mu)},
\label{5.3c}
\end{equation}
3 equations with 3 variables $T,\mu,V_{a}$;

the main process: no total confinement,
\begin{equation}
E(T,\mu)+\tilde{E}_{r}=E_{\Sigma},
\label{5.4a}
\end{equation}
\begin{equation}
N(T,\mu)+\tilde{N}_{r}=N_{\Sigma},
\label{5.4b}
\end{equation}
\begin{equation}
T=\frac{1}{8\pi\kappa E(T,\mu)},
\label{5.4c}
\end{equation}
3 equations with 4 variables $T,\mu,\tilde{E}_{r},
\tilde{N}_{r}$, one degree of freedom.

The main process in more detail:

a process: $N$-confinement,
\begin{equation}
E(T,\mu)+E_{a}(V_{a},T,\mu)+E_{r}=E_{\Sigma},
\label{5.5a}
\end{equation}
\begin{equation}
N(T,\mu)+N_{a}(V_{a},T,\mu)=N_{\Sigma},
\label{5.5b}
\end{equation}
\begin{equation}
T=\frac{1}{8\pi\kappa E(T,\mu)},
\label{5.5c}
\end{equation}
3 equations with 4 variables $T,\mu,V_{a},E_{r}$;

a process: no confinement,
\begin{equation}
E(T,\mu)+E_{a}(V_{a},T,\mu)+E_{r}=E_{\Sigma},
\label{5.6a}
\end{equation}
\begin{equation}
N(T,\mu)+N_{a}(V_{a},T,\mu)+N_{r}=N_{\Sigma},
\label{5.6b}
\end{equation}
\begin{equation}
T=\frac{1}{8\pi\kappa E(T,\mu)},
\label{5.6c}
\end{equation}
3 equations with 5 variables $T,\mu,V_{a},E_{r},N_{r}$;

a process: no atmosphere,
\begin{equation}
E(T,\mu)+E_{r}=E_{\Sigma},
\label{5.7a}
\end{equation}
\begin{equation}
N(T,\mu)+N_{r}=N_{\Sigma},
\label{5.7b}
\end{equation}
\begin{equation}
T=\frac{1}{8\pi\kappa E(T,\mu)},
\label{5.7c}
\end{equation}
3 equations with 4 variables $T,\mu,E_{r},N_{r}$;

a state: no atmosphere and $N$-exhaustion of the black hole,
\begin{equation}
E(T_{max},-\infty)+E_{r}=E_{\Sigma},
\label{5.8a}
\end{equation}
\begin{equation}
N_{r}=N_{\Sigma},
\label{5.8b}
\end{equation}
\begin{equation}
T_{max}=\frac{1}{8\pi\kappa E(T_{max},-\infty)},
\label{5.8c}
\end{equation}
3 equations with 3 variables $T_{max},E_{r},N_{r}$,
\begin{equation}
T_{max}\approx E(T_{max},-\infty)=\omega_{last}\approx m_{P};
\label{5.8d}
\end{equation}

a final process: no atmosphere, $N$-exhaustion of the black
hole, and no equilibrium of the black hole,
\begin{equation}
E(T,-\infty)+E_{r}=E_{\Sigma},
\label{5.9a}
\end{equation}
\begin{equation}
N_{r}=N_{\Sigma},
\label{5.9b}
\end{equation}
2 equation with 3 variables $T,E_{r},N_{r},\quad T<T_{max}$,
the black hole spits out the last quantum $\omega_{last}
\approx m_{P}$;

a final state: no black hole, no atmosphere,
\begin{equation}
E_{r}=E_{\Sigma},
\label{5.10a}
\end{equation}
\begin{equation}
N_{r}=N_{\Sigma},
\label{5.10b}
\end{equation}
2 equations with 2 variables $E_{r},N_{r}$, the history
of the black hole is finished.

In conclusion, it may be noted that it is possible to
measure the quantities $T,\left\{ \mu_{\alpha} \right\}$ for
the atmosphere and, by the same token, for the black hole.
Measuring $E$ through
\begin{equation}
E=\frac{R_{S}}{2\kappa}
\label{5.11}
\end{equation}
or
\begin{equation}
E=E_{\Sigma}-\tilde{E}_{r},
\label{5.12}
\end{equation}
we obtain
\begin{equation}
E=E(T,\left\{ \mu_{\alpha} \right\}).
\label{5.13}
\end{equation}

\section*{Acknowledgment}

I would like to thank Stefan V. Mashkevich for helpful
discussions.

\end{document}